\documentclass[12pt]{iopart}

\usepackage{bm}
\usepackage{epsfig}
\usepackage{citesort}
\usepackage{graphicx}
\eqnobysec

\newcommand{\lwig}{\mbox{\;\raisebox{.3ex}
    {$<$}$\!\!\!\!\!$\raisebox{-.9ex}{$\sim$}\;}}
\newcommand{\gwig}{\mbox{\;\raisebox{.3ex}
    {$>$}$\!\!\!\!\!$\raisebox{-.9ex}{$\sim$}}\;}
\newcommand{\lambdabar}%
{{\hbox{$\lambda$\kern-1.ex\raise+0.45ex\hbox{--}}}}

\begin{document}

%%%%%%%%%%%%%%%%%%%%%%%%%%%%%%%%%%%%%%%%%%%%%%%%%%%%%%%%%%%%%%%%%%%%%%
% Frontpage %%%%%%%%%%%%%%%%%%%%%%%%%%%%%%%%%%%%%%%%%%%%%%%%%%%%%%%%%%
%%%%%%%%%%%%%%%%%%%%%%%%%%%%%%%%%%%%%%%%%%%%%%%%%%%%%%%%%%%%%%%%%%%%%%

\begin{flushright}
{CERN-PH-TH/2009-044\\
LAPTH-1323/09\\
MPP-2009-37}
\end{flushright}

\title{Isocurvature forecast in the anthropic axion window}

\author{J.~Hamann$^1$, S.~Hannestad$^2$,
G.~G.~Raffelt$^3$ and Y.~Y.~Y.~Wong$^4$}

\address{$^1$~LAPTh, Universit\'e de Savoie,
CNRS  \\ BP 110, F-74941
Annecy-le-Vieux Cedex, France
\\
 $^2$~Department of Physics and Astronomy\\
 University of Aarhus, DK-8000 Aarhus C, Denmark\\
 $^3$~Max-Planck-Institut f\"ur Physik (Werner-Heisenberg-Institut)\\
 F\"ohringer Ring 6, D-80805 M\"unchen, Germany\\
$^4$~Theory Division, Physics Department, CERN,
CH-1211 Gen\`eve 23, Switzerland}

\ead{\mailto{hamann@lapp.in2p3.fr},
     \mailto{sth@phys.au.dk},
     \mailto{raffelt@mppmu.mpg.de} and  \\
     \mailto{yvonne.wong@cern.ch}}

\date{1 April 2009}

\begin{abstract}
We explore the cosmological sensitivity to the amplitude of
isocurvature fluctuations that would be caused by axions in the
``anthropic window'' where the axion decay constant $f_{\rm a}\gg
10^{12}$~GeV and the initial misalignment angle $\Theta_{\rm i}\ll
1$. In a minimal $\Lambda$CDM cosmology extended with subdominant
scale-invariant isocurvature fluctuations, existing data constrain the
isocurvature fraction to $\alpha<0.09$ at 95\% C.L. If no signal shows
up, Planck can improve this constraint to 0.042 while an ultimate CMB
probe limited only by cosmic variance in both temperature and
$E$-polarisation can reach 0.017, about a factor of five better than
the current limit.  In the parameter space of $f_{\rm a}$ and $H_I$
(Hubble parameter during inflation) we identify a small region where
axion detection remains within the reach of realistic cosmological
probes.
\end{abstract}

\maketitle

%%%%%%%%%%%%%%%%%%%%%%%%%%%%%%%%%%%%%%%%%%%%%%%%%%%%%%%%%%%%%%%%%%%%%%
\section{Introduction}                        \label{sec:introduction}
%%%%%%%%%%%%%%%%%%%%%%%%%%%%%%%%%%%%%%%%%%%%%%%%%%%%%%%%%%%%%%%%%%%%%%

The Peccei-Quinn (PQ) mechanism and concomitant axion is arguably the
most plausible explanation for the smallness of the $\Theta$
parameter of QCD~\cite{Peccei:2006as,Kim:2008hd}. This case has become
stronger in recent years because the experimental precision
exploration of the quark mixing matrix reveals that the
Kobayashi--Maskawa mechanism of explicit CP violation accounts
perfectly well for all observations~\cite{Amsler:2008zzb}, so one can
hardly argue that CP was a fundamental symmetry that is only
spontaneously broken. Likewise, the possibility of a massless up-quark
is now convincingly excluded~\cite{Amsler:2008zzb}, leaving few
credible alternatives to the PQ mechanism. Perhaps the greatest
advantage of the PQ mechanism is that it can be verified by the
detection of axions.

The CP-violating term in the QCD Lagrangian is of the form
$(\alpha_{\rm s}/8\pi)\Theta G_{\rm a}^{\mu\nu}\tilde
G_{a\mu\nu}$, with $\alpha_{\rm s}$ the strong fine structure constant
and $G$ the colour field-strength tensor. With the identification
$\Theta\to a/f_{\rm a}$ the axion field $a$ inherits its defining
interaction with the axial QCD anomaly. The axion decay constant
$f_{\rm a}$ is the primary parameter for axion physics. In particular,
all couplings are inversely proportional to $f_{\rm a}$ and the axion
mass is determined~by
\begin{equation}
\label{eq:axionmass}
 m_{\rm a} f_{\rm a}=\frac{\sqrt{m_{\rm u}m_{\rm d}}}{m_{\rm u}+m_{\rm d}}
 \,m_\pi f_\pi\, \equiv \Lambda_{\rm a}^2.
\end{equation}
Here $m_\pi=135.5$~MeV is the pion mass, $f_\pi=93$~MeV its decay
constant, and $m_{\rm u,d}$ the up- and down-quark masses with
$Z=m_{\rm u}/m_{\rm d}=0.35$--0.6 \cite{Amsler:2008zzb}. A series of
overlapping experimental and astrophysical constraints suggests
$f_{\rm a}\gwig 10^{9}$~GeV \cite{Amsler:2008zzb,Raffelt:2006cw}, so
axions are very light and ``invisible,'' in any case far more
elusive than neutrinos.

However, during the QCD epoch of the early universe, a non-thermal
mechanism produces axions as nonrelativistic coherent field
oscillations that can play the role of cold dark
matter~\cite{Sikivie:2006ni}. In terms of the initial ``misalignment
angle'' $\Theta_{\rm i}=a_{\rm i}/f_{\rm a}$ relative to the
CP-conserving minimum of the axion potential, the cosmic axion
density is~\cite{Bae:2008ue}
\begin{equation}
 \Omega_{\rm a}h^2\simeq 0.195\,\Theta_{\rm i}^2
 \left(\frac{f_{\rm a}}{10^{12}~{\rm GeV}}\right)^{1.184}\,.
\end{equation}
If $\Theta_{\rm i}^2$ is of order unity, axions
 provide the dark
matter of the universe if $f_{\rm a}\sim 10^{12}$~GeV
 ($m_{\rm
a}\sim10~\mu{\rm eV}$). As such they can be detected by
 experiments
that exploit the generic axion--two photon interaction,
 allowing for
the coherent conversion into excitations of a
 high-quality microwave
cavity placed in a strong
$B$-field~\cite{Asztalos:2006kz,Kim:2008hd}. If axions are in the
mass range 1--$100~\mu$eV and form the dark matter of our galaxy,
 the
ADMX and New CARRACK experiments are poised to find them within a
decade. Such a
 discovery would both reveal the nature of dark matter
and verify the
 PQ mechanism for CP symmetry restoration.\footnote{In
view of the
 fundamental importance of such a measurement it is
perplexing how
 little worldwide effort is directed towards this goal
compared with
 the vast range of WIMP dark matter searches.}
 
Meanwhile one may study other ``windows of opportunity'' where axions
could leave a detectable trace. Ongoing efforts include helioscope
searches that are sensitive to solar axions if $f_{\rm a}\sim
10^7$~GeV, corresponding to sub-eV
masses~\cite{Inoue:2008zp,Arik:2008mq}. Of course, if solar axions
were found something would have to be wrong with the supernova 1987A
energy-loss limit that requires $f_{\rm a}\gwig10^9$~GeV. For $f_{\rm
a}\lwig10^8$~GeV axions thermalise in the early universe after the
QCD epoch, so sub-eV mass axions would provide a hot dark matter
component~\cite{Hannestad:2005df,Melchiorri:2007cd,Hannestad:2008js}.
If future surveys were to reveal a hot dark matter fraction beyond
the
 minimal neutrino component, an axion interpretation would be
conceivable.
 
 We consider here axions in another range beyond the
classical
 cosmological window. In a scenario where the PQ symmetry is
not
 restored during or after inflation, a single value
$-\pi<\Theta_{\rm i}<+\pi$ determines the axion density. It is
possible that $\Theta_{\rm i}\ll1$, allowing for $f_{\rm a}\gg
10^{12}$~GeV. This case is motivated because the PQ mechanism
presumably is embedded in a greater framework. In particular, the PQ
symmetry emerges naturally in many string scenarios, but then it is
hard to obtain $f_{\rm a}$ much below
$10^{16}$~GeV~\cite{Svrcek:2006yi}.  The axion is the
Nambu--Goldstone
 boson of the global $U(1)_{\rm PQ}$ symmetry that is
broken at a scale
 $v_{\rm PQ}=N\,f_{\rm a}$ where $N$ is an
integer. (Both $a/v_{\rm
 PQ}$ and $a/f_{\rm a}$ must be
$2\pi$-periodic because $a/v_{\rm
 PQ}$ is the phase of a new Higgs
field responsible for spontaneous
 symmetry breaking and the QCD
vacuum is periodic in
 $\Theta=a/f_{\rm a}$.) Since $-\pi<a/v_{\rm
PQ}<+\pi$ encounters
 $N$ minima of the axion potential, there is a
formidable
 cosmological domain-wall problem unless $N=1$ or inflation
occurs
 after PQ symmetry breaking, further motivating a
``late-inflation
 scenario''~\cite{Sikivie:2006ni}.
 
 Postulating
$\Theta_{\rm i}\ll1$ may look like shuffling the QCD
 problem of a
small $\Theta$ parameter into an unnatural initial
 condition for our
universe. However, a GUT-scale value of $f_{\rm
 a}\sim 10^{16}$~GeV
would only require a modest fine-tuning of
 $\Theta_{\rm i}\lwig
3\times10^{-3}$ \cite{Pi:1984pv}. More
 importantly, anthropic
selection suggests that the cosmic
 baryon/dark matter ratio can not
be too small, so for any $f_{\rm
 a}$ the most probable value for
$\Theta_{\rm i}$ is near the one
 required for the observed
dark-matter
 density~\cite{Linde:1987bx,Tegmark:2005dy}. 
 
 Following
these papers we emphasise that in our late-inflation
 scenario
$\Theta_{\rm i}$, and therefore the axion dark matter density, is
 an
environmental parameter of our universe, not a fundamental one, and
that its prior probability distribution is known to be flat on the
interval $-\pi<\Theta_{\rm i}<+\pi$. This contrasts with typical other
cases
 of anthropic reasoning where usually it is not known if a
given
 cosmological or particle-physics parameter is fundamental or
environmental. In the axion case anthropic reasoning is unavoidable
to
 quantify if a given value of $\Theta_{\rm i}$ is natural or not.

%In contrast to typical
%cases of anthropic reasoning, we have here an unavoidable cosmic
%random process (the PQ symmetry breaking), so in our late-inflation
%scenario the axion density is a random number with a distribution
%given by the flat prior $-\pi<\Theta_{\rm
% i}<+\pi$. In other words, anthropic selection is unavoidable if one
%wishes to quantify the posterior probability distribution
%for~$\Theta_{\rm i}$.

Therefore, the ``anthropic axion window'' with $f_{\rm a}\gg
10^{12}$~GeV is well motivated. In this case $m_{\rm a}\ll 10~\mu$eV
and axion dark matter is difficult to detect with the cavity technique
because of the large required cavity size and magnetic field
region.\footnote{Ideas for overcoming this limitation were discussed
by S.~Thomas in a talk given at the workshop ``Axions at the Institute
for Advanced Study'' (20--22 Oct 2006, IAS, Princeton, New Jersey).}
An alternative signature is provided by primordial isocurvature
fluctuations that can show up in future data. In our late-inflation
scenario, the massless axion field is present during inflation and
thus acquires the usual amplitude fluctuations imprinted by the
de~Sitter expansion. When axions acquire a mass during the QCD epoch,
these fluctuations become dynamically relevant in the form of
isocurvature fluctuations that are uncorrelated with the adiabatic
fluctuations inherited by all other matter and radiation from the
inflaton field. The isocurvature amplitude depends on both $f_{\rm a}$
and $H_{\rm I}$, the Hubble parameter during inflation, so
observational limits on isocurvature fluctuations exclude certain
regions in this parameter space
\cite{Lyth:1989pb,Beltran:2006sq,Hertzberg:2008wr,Komatsu:2008hk,Kaplan:2008ss,Visinelli:2009zm}.

Since there is no trace of isocurvature fluctuations in existing
data, perhaps a more interesting question is the remaining window
for axions to show up in future. One forecast was recently provided
by the CMBPol Study Team Collaboration~\cite{Baumann:2008aq}.
The purpose of our paper is to
provide a more detailed sensitivity forecast on the amplitude of
isocurvature fluctuations.   We explore possible degeneracies that
may exist between the isocurvature fluctuations and other
cosmological parameters in both present and future data, and determine
the region in the underlying axion parameter space that is
realistically within the reach of future cosmological probes.

To this end we review in Sec.~\ref{sec:observables} the axion dark
matter and isocurvature contributions in terms of underlying
physical parameters. In Sec.~\ref{sec:constraints} we derive our
constraints and sensitivity forecasts on the axion-type isocurvature
fraction. We interpret these results in terms of underlying physical
parameters in Sec.~\ref{sec:axionparameters} and conclude in
Sec.~\ref{sec:conclusions}.

%%%%%%%%%%%%%%%%%%%%%%%%%%%%%%%%%%%%%%%%%%%%%%%%%%%%%%%%%%%%%%%%%%%%%
\section{Cosmological imprint of axions      \label{sec:observables}}
%%%%%%%%%%%%%%%%%%%%%%%%%%%%%%%%%%%%%%%%%%%%%%%%%%%%%%%%%%%%%%%%%%%%%

\subsection{Energy density}

To specify the cosmological imprint of axions we begin with their
contribution to the matter density of the present-day universe. When
the PQ symmetry breaks at some large temperature $T\sim v_{\rm PQ}$,
the relevant Higgs field will settle in a minimum corresponding to
$\Theta_{\rm i}=a_{\rm i}/f_{\rm a}$, where $-\pi\leq\Theta_{\rm
  i}\leq+\pi$.  We assume that this happens before cosmic inflation,
so throughout our observable universe we have the same initial
condition except for fluctuations imprinted by inflation itself. When
the universe cools and evolves towards the QCD epoch, instanton effects
become important and produce a potential for the axion field.  The
corresponding temperature-dependent axion mass is~\cite{Bae:2008ue}
\begin{equation}\label{eq:mass}
m_{\rm a}(T) \simeq  \left\{
\begin{array}{ll}
\beta_{\rm inst}^{1/2}\left(\frac{T}{\rm GeV}
\right)^{-n/2}\!\Big/f_{\rm a}
& \quad{\rm for}~ T \gwig \Lambda_{\rm QCD}, \\
\Lambda_{\rm a}^2/f_{\rm a}  & \quad{\rm for}~T \ll  \Lambda_{\rm QCD}, \\
\end{array} \right.
\end{equation}
where $\beta_{\rm inst} \simeq 3.964~{\rm MeV}^4$, $n \simeq 6.878$,
$\Lambda_{\rm QCD}\simeq 380~{\rm MeV}$, and $\Lambda_{\rm a} \simeq
77~{\rm MeV}$, assuming $Z=0.5$.

The axion field begins oscillating at $3 H(T_{\rm osc}) \simeq m_{\rm
  a}(T_{\rm osc})$.  The Hubble parameter during radiation domination
is
\begin{equation}\label{eq:hubble}
 H^2(T) = \frac{8 \pi}{3 M_{\rm Pl}^2}\,
 \frac{\pi^2}{30}\, g_*(T)\, T^4\,,
\end{equation}
where $M_{\rm Pl}=1.221 \times 10^{19} \ {\rm GeV}$ is the Planck
mass.  For $T \gwig \Lambda_{\rm QCD}$ the effective number of
thermally excited degrees of freedom is $g_*(T_{\rm osc}) \simeq
61.75$, whereas $g_*(T_{\rm osc}) \simeq 10.75$ for $T \ll
\Lambda_{\rm QCD}$.  Together equations~(\ref{eq:mass}) and
(\ref{eq:hubble}) give
\begin{equation}
T_{\rm osc} \simeq \left\{
\begin{array}{ll}
9.16 \times 10^{2} \ {\rm MeV}
\left( \frac{f_{\rm a}}{10^{12} \ {\rm GeV}} \right)^{-0.184}
&\quad{\rm for}~T_{\rm osc} \gwig \Lambda_{\rm QCD},\\
6.65 \times 10^4 \ {\rm MeV}
\left( \frac{f_{\rm a}}{10^{12} \ {\rm GeV}} \right)^{-0.5}
&\quad{\rm for}~T_{\rm osc} \ll \Lambda_{\rm QCD}.
\end{array} \right.
\end{equation}
We see that the limit $T_{\rm osc}\ll\Lambda_{\rm QCD}$ is only
relevant for $f_{\rm a} \gg 3 \times 10^{16}~{\rm GeV}$. We will
usually focus on $f_{\rm a}$ values around or below $10^{16}~{\rm
  GeV}$ and therefore concentrate on the case where the axion field
begins oscillating during the QCD epoch.

The traditional calculation of the present-day axion
density~\cite{Sikivie:2006ni} was recently revisited in detail, taking
into account modern values for the relevant QCD parameters and
anharmonic corrections for the axion
potential~\cite{Bae:2008ue}. However, since we are interested in the
anthropic axion window where the initial misalignment angle is small,
anharmonic effects are not important. We also ignore the possibility
of axion dilution by late entropy production, so
\begin{equation}\label{eq:omc}
\omega_{\rm a} \equiv \Omega_{\rm a} h^2 \simeq
0.195  \left(\frac{f_{\rm a}}{10^{12} \ {\rm GeV}}\right)^{1.184}
 (\Theta_{\rm i}^2 + \sigma_\Theta^2) \,.
\end{equation}
Here, the initial misalignment angle $\Theta_{\rm i}$ is to be
interpreted as the average in our Hubble volume, $\Theta_{\rm
  i}=\langle\Theta\rangle_{\rm i}$, while
\begin{equation}
\label{eq:sigma}
\sigma_\Theta^2 = \frac{H_I^2}{4 \pi^2 f_{\rm a}^2}
\end{equation}
is the inflation-induced variance, with $H_I$ the Hubble parameter
during inflation. In other words, the relevant initial parameter is
$\langle \Theta^2\rangle_{\rm i}=\Theta_{\rm i}^2+\sigma_\Theta^2$.

We will usually assume that all of the cold dark matter consists of
axions, so according to current cosmological data $\omega_{\rm
  a}=\omega_{\rm c} \simeq 0.11$ \cite{Komatsu:2008hk}. 
  Assuming $\sigma_\Theta$ is small, we then find
\begin{equation}
\Theta_{\rm i} \simeq 0.75\,
 \left(\frac{10^{12}~{\rm GeV}}{f_{\rm a}}\right)^{0.592}
\end{equation}
as a unique relationship between the initial misalignment angle and
the axion decay constant.

\subsection{Isocurvature fraction}

Since the axion is essentially massless during inflation, quantum
fluctuations are imprinted on the axion field in the same manner
that the inflaton field acquires its perturbations.  Because of the
axion's negligible contribution to the total energy budget of the
universe during inflation, these fluctuations do not perturb the
total energy density and manifest themselves only as perturbations
in the ratio of axion number density to entropy density, i.e.,
$\delta \rho = 0$, and $\delta (n_{\rm a}/s) \neq 0$.  For this
reason they are known as entropy or isocurvature perturbations.  As
the axion field provides some or all of the cold dark matter of the
universe, these perturbations contribute to the temperature and
polarisation fluctuations in the CMB.

Axion-induced isocurvature fluctuations are uncorrelated with the
adiabatic fluctuations inherited by other matter and radiation
components from the inflaton. The spectrum of entropy
perturbations~is
\begin{equation}
{\cal S}(k)=
\frac{\Theta^2- \langle\Theta^2\rangle}{\langle\Theta^2\rangle}.
\end{equation}
Assuming $\delta \Theta \equiv \Theta-\langle \Theta \rangle$ has a
Gaussian distribution, the perturbation power spectrum can be
written~as
\begin{equation}
\langle |{\cal S}(k)^2| \rangle =
\frac{2 \sigma^2_\Theta\,(2 \Theta_{\rm i}^2 + \sigma^2_\Theta)}
{(\Theta_{\rm i}^2 + \sigma^2_\Theta)^2},
\end{equation}
where $\Theta_{\rm i}$ and $\sigma_\Theta^2$ are the mean and the
variance of the initial $\Theta$ distribution.  As we shall see
later, for all of the viable parameter space, $\Theta_{\rm i}^2 \gg
\sigma_\Theta^2$, thus
\begin{equation}
 \langle |{\cal S}(k)^2| \rangle \simeq \frac{H_I^2}
 {\pi^2 f_{\rm a}^2 \Theta_{\rm i}^2}
 \propto \left( \frac{k}{k_0} \right)^{n_{\rm iso}-1}\,.
\end{equation}
In the second equality we have approximated the power spectrum by a
power law with spectral index $n_{\rm iso}=1-2 \epsilon \simeq 1$,
where $\epsilon$ is the first inflationary slow-row roll parameter
and $k_0=0.002~{\rm Mpc}^{-1}$ the pivot scale. Likewise, adiabatic
perturbations sourced by the inflaton field are encoded in the
curvature power spectrum,
\begin{equation}
{\langle |{\cal R}(k)^2| \rangle}
= \frac{H_I^2}{\pi M_{\rm Pl}^2 \epsilon}
\propto \left( \frac{k}{k_0} \right)^{n_{\rm ad}-1},
\end{equation}
where $n_{\rm ad}$ is the adiabatic spectral index.

Because fluctuations of the axion field are uncorrelated with those of
the inflaton field, the primordial scalar power spectrum is simply the
incoherent sum of the adiabatic and isocurvature perturbation spectra,
\begin{equation}
{\cal P}(k) =
\langle |{\cal R}(k)^2| \rangle+\langle |{\cal S}(k)^2| \rangle\,.
\end{equation}
The resulting CMB temperature and polarisation anisotropy spectra are
given by a similar sum, but with the appropriate transfer functions
$T_{X}(k)$ folded in ($X=T,E$),
\begin{eqnarray}
C_{XY}^{\rm total}(\ell)  &=&
C_{XY}^{\rm ad} (\ell)+ C_{XY}^{\rm iso}(\ell) \nonumber \\
& \propto & \int \frac{dk}{k}\ T^{\rm ad}_X(k) T^{\rm ad}_Y(k) {\langle |{\cal R}(k)^2| \rangle}+
\int \frac{dk}{k} \ T^{\rm iso}_X(k) T^{\rm iso}_Y(k) {\langle |{\cal S}(k)^2| \rangle}.\nonumber \\
\end{eqnarray}
Here it is useful to define the isocurvature fraction $\alpha$,
\begin{equation}
\alpha \equiv \left. \frac{\langle |{\cal S}(k)^2| \rangle}
{\langle |{\cal R}(k)^2| \rangle+\langle |{\cal S}(k)^2| \rangle } \right|_{k=k_0}
\simeq \left\{ \begin{array}{ll}
\frac{H_I^2}
{A_{\rm S} \pi^2 f_{\rm a}^2 \Theta_{\rm i}^2}
&\quad{\rm for}~\Theta_{\rm i}^2 \gg \sigma_\Theta^2, \\
\frac{2}{A_{\rm S}}
&\quad{\rm for}~\Theta_{\rm i}^2 \ll \sigma_\Theta^2, \\
\end{array} \right.
\end{equation}
where $A_{\rm S}= {\cal P}(k=k_0)$ is the amplitude of the total
primordial scalar power spectrum at the pivot scale $k_0$, and we have
assumed $\alpha \ll 1$.  Given $A_{\rm S} \simeq 10^{-9}$, the
solution $\Theta_{\rm i}^2 \ll
\sigma_\Theta^2$ cannot be realised.
In the opposite limit $\Theta_{\rm i}^2 \gg \sigma_\Theta^2$ we find
\begin{eqnarray}\label{eq:axionalpha}
\alpha &\simeq&7.5\times10^{-3}
\left(\frac{2.4 \times 10^{-9}}{A_{\rm S}} \right)
\left(\frac{0.109}{\omega_{\rm c}} \right) \left(
\frac{H_I}{10^{7} \ {\rm GeV}}\right)^2  \left( \frac{10^{12} \ {\rm
GeV}}{f_{\rm a}} \right)^{0.816},\nonumber\\
\end{eqnarray}
where we have used equation~(\ref{eq:omc}).

\subsection{Gravity waves}

The misalignment mechanism {\it per se\/} does not produce gravity
waves. However, all inflationary scenarios produce some amount of
tensor perturbations, which is in principle manifest in the CMB
temperature and polarisation anisotropies.  The quantity of interest
is the tensor-to-scalar ratio $r=A_{\rm T}/A_{\rm S}$, defined as
the ratio of the primordial tensor to the scalar perturbation spectrum
at $k=k_0$. The amplitude of tensor perturbations from inflation is
$A_{\rm T} = 16 H_I^2/(\pi M_{\rm Pl}^2)$.  Thus, the Hubble
expansion rate during inflation
\begin{eqnarray}
H_I &=& \frac{1}{4} \sqrt{\pi A_{\rm S} r} \ M_{\rm Pl}\nonumber \\ &
\simeq& 1.33 \times 10^{14} \ {\rm GeV} \left(\frac{A_{\rm S}}{2.4
\times 10^{-9}} \right)^{1/2}
\left( \frac{r}{0.25} \right)^{1/2}
\end{eqnarray}
is directly probed by a measurement of $r$.

%%%%%%%%%%%%%%%%%%%%%%%%%%%%%%%%%%%%%%%%%%%%%%%%%%%%%%%%%%%%%%%%%%%%%%%
\section{Present constraints and future sensitivities
\label{sec:constraints}}
%%%%%%%%%%%%%%%%%%%%%%%%%%%%%%%%%%%%%%%%%%%%%%%%%%%%%%%%%%%%%%%%%%%%%%%

\subsection{Observational signatures of axion-type isocurvature
perturbations}

\begin{figure}
\begin{indented}
%\item\includegraphics[height=374pt, angle=270]{fig1.eps}
\item\includegraphics[height=340pt, angle=270]{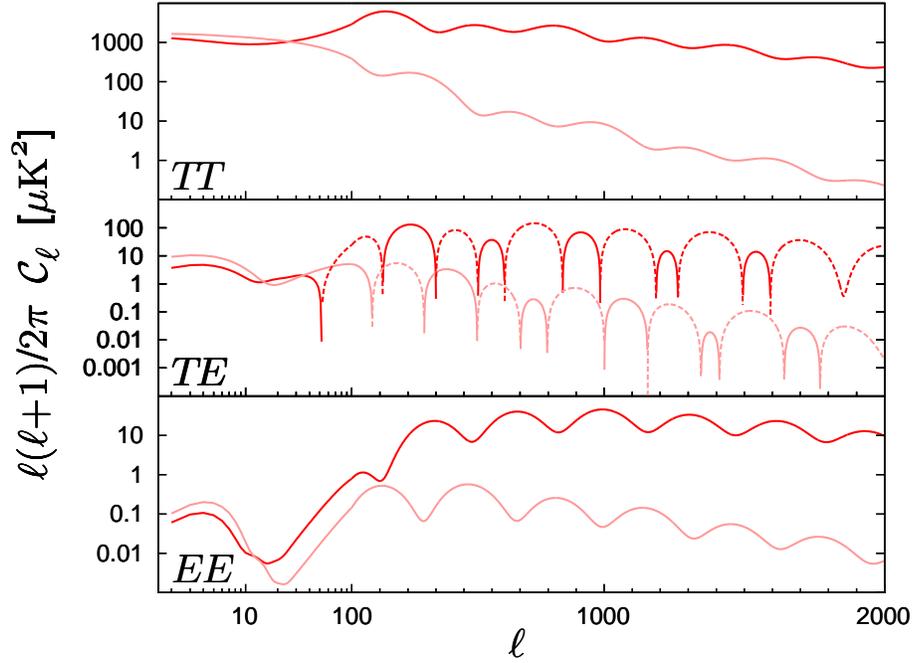}
\end{indented}
\caption{CMB angular power spectra for purely adiabatic initial
conditions ($\alpha = 0$, dark red lines) with $n_{\rm ad} = 0.96$ and
for purely CDM isocurvature initial conditions ($\alpha = 1$, light
red lines) with a scale invariant spectrum $n_{\rm iso} = 1$. All
other cosmological parameters are kept fixed. \label{fig:isocls}}
\end{figure}

To discuss the observational imprint in the CMB we recall that the
axion field causes scale-invariant CDM isocurvature fluctuations
that are uncorrelated with the dominant inflaton-induced adiabatic
fluctuations. Compared to adiabatic initial conditions, the growth
of isocurvature CDM density perturbations with wavenumber $k$ is
suppressed by a factor of $k^2$ on sub-horizon scales
\cite{Bucher:1999re}.  Hence, if we are looking for traces of a
subdominant isocurvature perturbation with a similar primordial
scale dependence as the dominant adiabatic mode, the signal will be
strongest at large scales.  We illustrate this point in
Fig.~\ref{fig:isocls}, where we plot the CMB angular power spectra
for a model with purely adiabatic perturbations ($\alpha = 0$) and
$n_{\rm
  ad} = 0.96$, and for a corresponding model with purely CDM
isocurvature perturbations ($\alpha = 1$) with $n_{\rm iso} = 1$.  For
uncorrelated isocurvature perturbations, cases with $0 < \alpha < 1$
simply correspond to weighted sums of these two spectra, so even for
largish values of $\alpha$, the spectrum at multipoles $\ell \gwig
200$ will be completely dominated by the adiabatic contribution.

\begin{figure}[ht]
\begin{indented}
%\item\includegraphics[height=374pt, angle=270]{fig2.eps}
\item\includegraphics[height=340pt, angle=270]{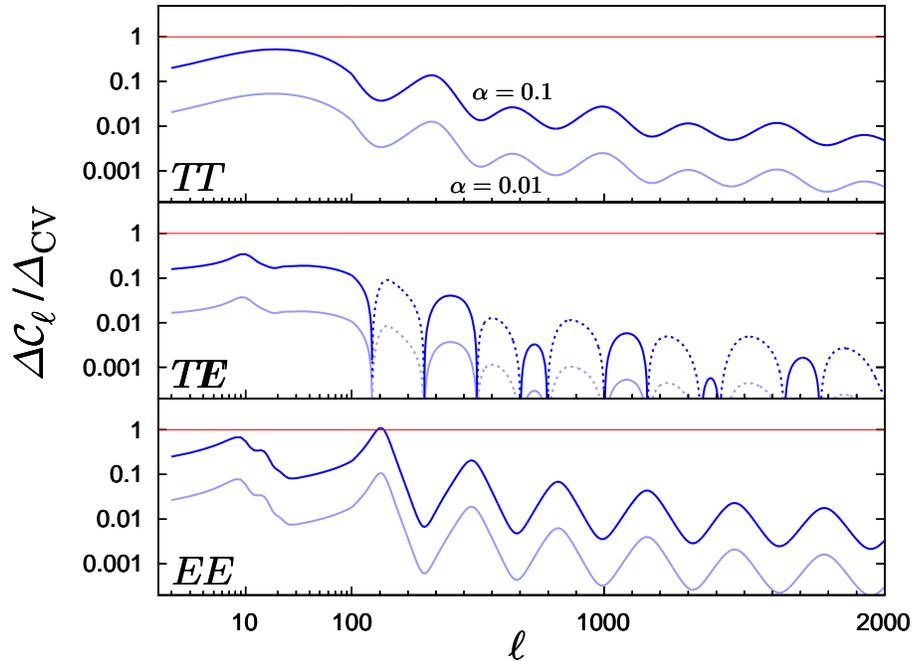}
\end{indented}
\caption{Scale dependence of the
``signal to noise'' for isocurvature fractions of $\alpha = 0.1$ (blue
lines), and $\alpha = 0.001$ (pale blue lines). Plotted is the
distortion $\Delta \mathcal{C}_\ell$ of the CMB angular power spectra
caused by adding a subdominant scale-invariant uncorrelated CDM
isocurvature component, given in units of the cosmic variance
$\Delta_{\rm CV}$. \label{fig:isoclcv}}
\end{figure}

It is also useful to contrast the isocurvature signal with the
cosmic variance, which poses a fundamental limit to the amount of
information to be extracted from CMB anisotropies.  In
Fig.~\ref{fig:isoclcv} we compare the contribution $\Delta
\mathcal{C}_\ell$ of a subdominant isocurvature part to the total
$\mathcal{C}_\ell$ with the intrinsic uncertainty due to cosmic
variance, $\Delta_{\rm CV}^{XY} = \sqrt{\mathcal{C}_\ell^{XX}
\mathcal{C}_\ell^{YY}} \sqrt{\frac{2}{2\ell+1}}$, with $X,Y \in
[T,E]$.  This shows that data of small-scale perturbations by
themselves carry hardly any information about axion-type
isocurvature.

Based on existing measurements of the CMB anisotropies, the main
signature of scale-independent uncorrelated CDM isocurvature, i.e., an
enhancement of the Sachs--Wolfe plateau, can be mimicked to a certain
extent by changing the values of other cosmological parameters.  In
particular, one can expect $\alpha$ to be degenerate with the total
matter density $\Omega_{\rm m}$ and the adiabatic spectral index
$n_{\rm ad}$.  Adding data of the large-scale matter power spectrum
and of distance probes such as supernovae will however allow us to
break these degeneracies and thus improve constraints on $\alpha$
indirectly.

\subsection{Constraints and forecast}

For inferring constraints from present data and predicting the
discovery potential of future CMB experiments, we consider a
cosmological model with seven free parameters: the isocurvature
fraction $\alpha$ plus the usual six parameters of the ``vanilla''
$\Lambda$CDM model. These parameters and their benchmark values to
generate mock data for our forecasts are the baryon density
$\omega_{\rm b} = 0.0226$, the dark matter density $\omega_{\rm c} =
0.109$, the Hubble parameter $h=0.715$, the redshift to reionisation
$z_{\rm re}=11.36$, the scalar spectral index $n_{\rm ad}=0.962$,
the normalisation $\ln( 10^{10} A_{\rm S}) = 3.195$, and the
isocurvature fraction $\alpha=0$.  The spectral index of the
primordial isocurvature power spectrum is fixed to $n_{\rm iso}=1$.
For the other parameters we adopt flat priors in the data analysis.

We do not vary the tensor-to-scalar ratio $r$, since in the anthropic
parameter region any possibly observable amount of tensor
perturbations would imply a value of $H_I$ high enough to make the
isocurvature fraction close to unity, which is clearly inconsistent
with observations~\cite{Fox:2004kb}.

Parameters are estimated using standard Markov Chain Monte Carlo
techniques, as implemented in the publicly available package
\texttt{CosmoMC}~\cite{Lewis:2002ah}. As input we use four different
data sets:
\begin{itemize}
\item[1.]{WMAP: the five-year WMAP data \cite{Nolta:2008ih}.}
\item[2.]{WMAP+SDSS-LRG+SN: WMAP plus the galaxy power spectrum
    from the luminous red galaxy subsample of the Sloan Digital
    Sky Survey~\cite{Tegmark:2006az} and the luminosity
    distances of supernovae from the Union
    compilation~\cite{Kowalski:2008ez}.}
\item[3.]{Planck: Simulated $TT$, $TE$ and $EE$ spectra up to
    $\ell = 2000$ from the Planck satellite \cite{planck},
    assuming isotropic white noise, a sky coverage of 70\%, and
    14 months of observation in the 70, 100, and 143~GHz
    channels.}
\item[4.]{CVL: Simulated, noiseless (i.e., cosmic variance
    limited) $TT$, $TE$ and $EE$ spectra up to $\ell = 2000$.}
\end{itemize}
The mock CMB data sets 3 and 4 are generated using the method
discussed in Ref.~\cite{Perotto:2006rj}. The likelihood
function ${\cal L}$ is defined as
\begin{equation}
\label{eq:likelihood}
\chi^2_{\rm eff} \equiv -2 \ln {\cal L} = \sum_{\ell=2}^{\ell_{\rm
max}} (2 \ell +1) \ f_{\rm sky} \left[{\rm Tr}(\widetilde{\bm
C}_\ell^{-1} \hat{\bm C}_\ell) + \ln \frac{|\widetilde{\bm
C}_\ell|}{|\hat{\bm C}_\ell|} - n \right],
\end{equation}
where $\hat{\bm C}_\ell \doteq \hat{C}^{XY}_\ell$ with $X,Y \in
[T,E]$ denotes the mock data covariance matrix, $\widetilde{\bm
C}_\ell \doteq C^{XY}_\ell + N^{XY}_\ell$ is the total covariance
matrix comprising theoretical predictions of the CMB anisotropy
spectrum $C^{XY}_\ell$ and the noise power spectrum $N^{XY}_\ell$.
The quantity $n$ counts the number of observable modes, where $n=2$
for observations in temperature and $E$-type polarisation.

\begin{figure}[t]
\begin{indented}
\item\includegraphics[height=340pt, angle=270]{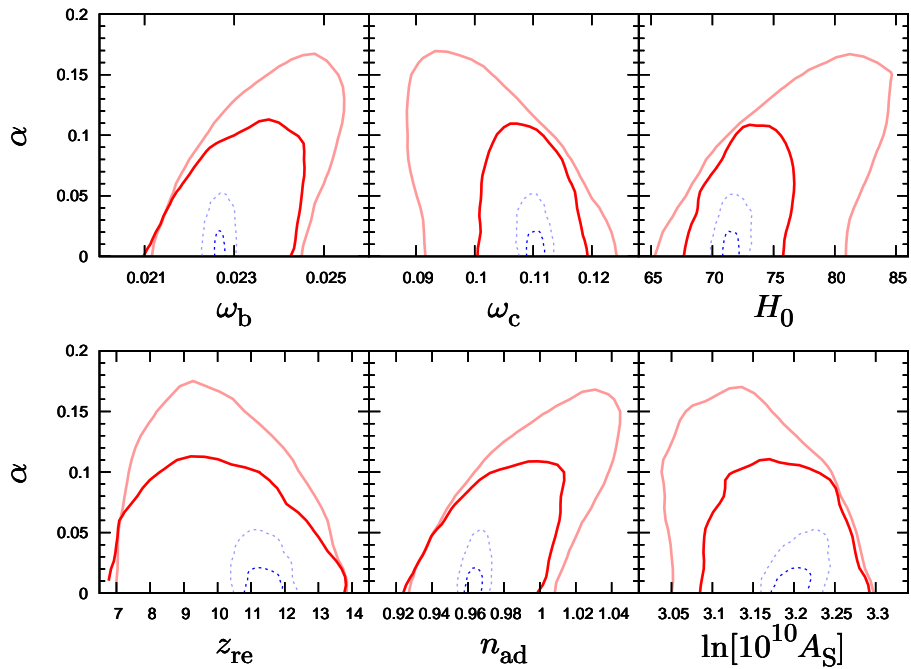}
\end{indented}
\caption{Degeneracies of the isocurvature fraction~$\alpha$ with the
other parameters of the standard $\Lambda$CDM-model for different data
sets.  Plotted are the marginalised two-dimensional 95\% credible
contours for WMAP data only (pale red), WMAP+SDSS-LRG+SN (red), and
for simulated data from Planck (thin pale blue dotted) and a cosmic
variance limited experiment (thin blue dotted lines). \label{fig:2d}}
\end{figure}

Using data set 2, we find 95\% credible intervals for the
observables discussed in Sec.~\ref{sec:observables},
\begin{eqnarray}\label{eq:currentconstraint}
0.1023 &<& \omega_{\rm c}  < 0.1165,  \\
3.105 &<& \ln [10^{10} A_{\rm S}] < 3.260, \\
\quad \alpha &<& 0.09,
\end{eqnarray}
consistent with the findings of Komatsu et
al.~\cite{Komatsu:2008hk}. As noted by these authors, adding
large-scale structure and supernova data leads to a significantly
tighter bound on $\alpha$ because these data sets contain additional
information on $\Omega_{\rm m}$, and thus help alleviate the
degeneracy problem somewhat.  This is illustrated in
Fig.~\ref{fig:2d}, where we show the two-dimensional joint
constraints on $\alpha$ with the other parameters of the vanilla
model for data sets 1--4.

For the future experiments and if no isocurvature signal shows up,
we forecast 95\%-credible upper limits of
\begin{eqnarray}\label{eq:sensitivities}
\alpha &< 0.042 \quad \quad &{\rm (Planck)},\\
\alpha &< 0.017 \quad \quad & \;{\rm (CVL)},
\end{eqnarray}
in rough agreement with the CMBpol White
Paper~\cite{Baumann:2008aq}.%
\footnote{The
noise properties of the CMB probes considered in
Ref.~\cite{Baumann:2008aq} are roughly equivalent to our CVL
experiment. Some other subtle differences include a slightly larger
sky coverage of 80\% (which leads to a general $\sqrt{0.8/0.7}-1 \sim
7$\% improvement in the parameter sensitivities compared with numbers
derived from a 70\% sky coverage) and the inclusion of
$B$-polarisation (which has no bearing on isocurvature detection).  A
major difference between Ref.~\cite{Baumann:2008aq} and our analysis
is the forecast method: we have analysed mock future data, while the
forecasts of Ref.~\cite{Baumann:2008aq} are based on the Fisher matrix
which assumes Gaussian posterior distributions.}
We emphasise that upcoming CMB experiments will break essentially all
remaining degeneracies without the need for additional data. The
improvement one can expect from Planck is partly due to this effect,
but the improved sensitivity to $E$-polarisation will also play a role
(Planck $E$-polarisation data will essentially be cosmic variance
limited for $\ell \lwig \mathcal{O}(10)$). Going from Planck to CVL
will increase the sensitivity to $\alpha$ by another factor of 2.5,
mostly because CVL covers the ``sweet spot'' in the $E$-polarisation
signal-to-noise, where the first acoustic peak of the isocurvature
signal coincides with the first acoustic trough of the adiabatic
signal, around $\ell=200$ (see Fig.~\ref{fig:isoclcv}). Because the
isocurvature signal affects predominantly the largest scales, any
significant further increase in the sensitivity to $\alpha$ from CMB
observations beyond the CVL benchmark value is most likely
unrealistic.

%%%%%%%%%%%%%%%%%%%%%%%%%%%%%%%%%%%%%%%%%%%%%%%%%%%%%%%%%%%%%%%%%%%%%%
\section{Axion parameter space            \label{sec:axionparameters}}
%%%%%%%%%%%%%%%%%%%%%%%%%%%%%%%%%%%%%%%%%%%%%%%%%%%%%%%%%%%%%%%%%%%%%%

\begin{figure}
\begin{indented}
\item\includegraphics[width=374pt]{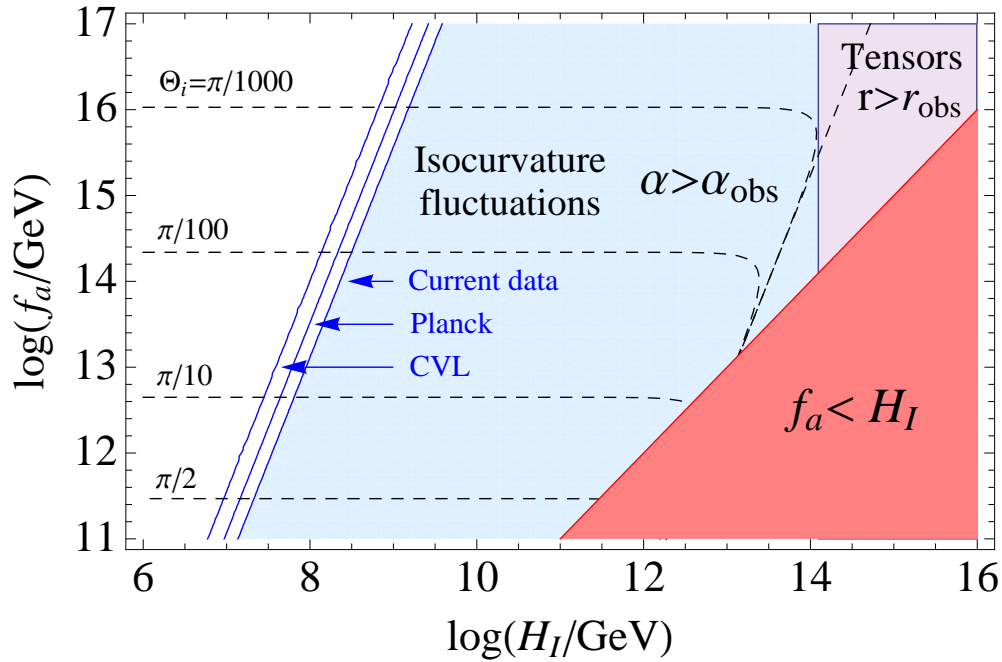}
\end{indented}
\caption{Exclusion and sensitivity regions in the plane of $H_I$
(Hubble rate during inflation) and $f_{\rm a}$ (axion decay
constant), assuming axions are all of the dark matter. The
isocurvature exclusion region based on current data is shown in
light blue with a boundary given by
equation~(\ref{eq:currentconstraint}). The sensitivity forecasts for
Planck and CVL of equation~(\ref{eq:sensitivities}) are also
indicated. The dashed lines indicate the required $\Theta_{\rm i}$
for a given $f_{\rm a}$
to obtain the full amount of axion dark matter. We also show the
region of excessive tensor modes and the region $f_{\rm a}<H_I$
where our late-inflation scenario is not applicable.\label{fig:tegmark}}
\end{figure}

It is now easy to translate our constraints and sensitivity
forecasts on the isocurvature fraction $\alpha$ into axion
parameters by virtue of equation~(\ref{eq:axionalpha}). It can be
written~as
\begin{equation}
H_I=3.5\times10^7~{\rm GeV}
\left(\frac{\alpha}{0.09}\right)^{1/2}
\left(\frac{\omega_{\rm c}}{0.109}\right)^{1/2}
\left(\frac{f_{\rm a}}{10^{12}~{\rm GeV}}\right)^{0.408},
\end{equation}
where we have used our present-day $\alpha$-constraint from
equation~(\ref{eq:currentconstraint}) as a benchmark value. Assuming
axions are the dark matter, we show this constraint in
Fig.~\ref{fig:tegmark} with a line marked ``Current data.''

In this plot, which is inspired by a similar figure in
Ref.~\cite{Hertzberg:2008wr}, we also show the relationship between
$f_{\rm a}$ and $\Theta_{\rm i}$ as dashed lines marked with the
relevant value of $\Theta_{\rm i}$. We have also marked the region
where the PQ scale is smaller than $H_I$ and where therefore our
``late inflation scenario'' would not apply. Also displayed is the
region of $H_I$ values excluded by excessive tensor modes,
corresponding to $r > 0.25$. Future sensitivities to $\alpha$ from
Planck and CVL based on equation~(\ref{eq:sensitivities}) are shown
labelled ``Planck'' and ``CVL'' respectively.

%%%%%%%%%%%%%%%%%%%%%%%%%%%%%%%%%%%%%%%%%%%%%%%%%%%%%%%%%%%%%%%%%%%%%%
\section{Conclusions                          \label{sec:conclusions}}
%%%%%%%%%%%%%%%%%%%%%%%%%%%%%%%%%%%%%%%%%%%%%%%%%%%%%%%%%%%%%%%%%%%%%%

We have studied current cosmological limits and future sensitivities
for axion-type isocurvature fluctuations. Our present-day limit on
the isocurvature fraction $\alpha<0.09$ (95\% C.L.) agrees with
similar recent results by other authors. We have used a $\Lambda$CDM
cosmological model with six standard parameters and in addition a
scale-invariant isocurvature fluctuation spectrum that is
uncorrelated with the dominant adiabatic fluctuations.

The main effect of axion isocurvature fluctuations is to modify the
Sachs--Wolfe plateau at small CMB multipole orders.\footnote{This
phenomenological limitation can be circumvented in a class of extended
axion models that lead to a significantly blue-tilted spectrum
\cite{Kasuya:2009up}, affecting also CMB anisotropies at larger
multipoles.} However, breaking degeneracies with other parameters,
future CMB probes can significantly improve the sensitivity to
$\alpha$ and thus to the presence of axions in the universe. A
non-detection of isocurvature modes by Planck would improve the limit
to $\alpha<0.042$, while an ultimate CMB probe limited only by cosmic
variance in both temperature and $E$-polarisation up to $\ell =2000$
could reach $\alpha<0.017$, about a factor of five more restrictive
than current limits.  These forecasts agree roughly with previous
estimates, notably by the CMBPol
Collaboration~\cite{Baumann:2008aq}. The small differences are
probably due to their using Fisher-matrix techniques in contrast to
our full-fledged analysis of mock data sets.

We emphasise that the CVL sensitivity to $\alpha$ is essentially ``as
good as it gets'', since all relevant degeneracies have been lifted
and no other cosmological probe besides the CMB is directly sensitive
to isocurvature fluctuations.  In terms of the axion decay constant
$f_{\rm a}$, the cosmic-variance limited probe shifts the sensitivity
region by approximately a factor of five for some fixed Hubble
parameter during inflation $H_I$, and thus opens a sliver of parameter
space where axions can still leave a detectable imprint in the
microwave sky. If a tentative signal were found, an experimental
search for axion dark matter in this parameter range would be crucial,
yet technologically challenging.

The idea of dark-matter axions with $f_{\rm a}$ at the GUT scale or
larger has received relatively little attention, at least from the
perspective of experimental searches, because the initial misalignment
angle would have to be much smaller than unity.  However, in our
late-inflation scenario the axion dark matter density is fixed by the
random number $-\pi<\Theta_{\rm i}<+\pi$ with a flat prior
distribution. In this case anthropic selection is practically
unavoidable and implies that the required small $\Theta_{\rm i}$ value
is not unnatural. Therefore, as stressed by previous authors, the
anthropic axion window is a plausible parameter range. In the absence
of direct experimental searches in this window, the appearance of
isocurvature fluctuations is the only opportunity for axions to become
visible in this scenario.  However, in view of the fundmental
limitations posed by cosmic variance on all cosmological probes, a
large chunk of parameter space in $f_{\rm a}$ and $H_I$ will likely
remain unreachable by this method.

%%%%%%%%%%%%%%%%%%%%%%%%%%%%%%%%%%%%%%%%%%%%%%%%%%%%%%%%%%%%%%%%%%%%%%
\section*{Acknowledgements}
%%%%%%%%%%%%%%%%%%%%%%%%%%%%%%%%%%%%%%%%%%%%%%%%%%%%%%%%%%%%%%%%%%%%%%

We acknowledge use of computing resources from the Danish Center for
Scientific Computing (DCSC). In Munich, partial support by the
Deutsche Forschungsgemeinschaft under the grant TR~27 ``Neutrinos and
beyond'' and the Cluster of Excellence ``Origin and Structure of the
Universe'' is acknowledged.

%%%%%%%%%%%%%%%%%%%%%%%%%%%%%%%%%%%%%%%%%%%%%%%%%%%%%%%%%%%%%%%%%%%%%%
\section*{References}
%%%%%%%%%%%%%%%%%%%%%%%%%%%%%%%%%%%%%%%%%%%%%%%%%%%%%%%%%%%%%%%%%%%%%%

\end{document}